\documentclass[conference]{IEEEtran}

\usepackage{cite}
\usepackage{url}
\usepackage{bbm}
\usepackage{amsmath,amsfonts} 
\usepackage{bbding}
\usepackage{float}
\usepackage{graphicx}
\usepackage{algorithm}
\usepackage{algpseudocode}
\usepackage{siunitx}
\usepackage[labelfont=bf]{caption}
\usepackage{subcaption}
\usepackage{tikz}
\usepackage{pgfplots}
\usepackage{epstopdf}
\usepackage{multirow}
\usepackage[font=scriptsize]{caption}
\usepackage{soul}
\usepackage{color, colortbl}
\usepackage{array}
\usepackage[T1]{fontenc}
\usepackage[utf8]{inputenc}
\usepackage{mathptmx}
\usepackage{listings}
\usepackage{adjustbox}
\usepackage{makecell}
\usepackage{tcolorbox}
\usepackage{threeparttable}
\usepackage[top=0.75in, left=0.65in, right=0.65in]{geometry} 

\usepackage{enumitem}
\usepackage{pifont}

\def\BibTeX{{\rm B\kern-.05em{\sc i\kern-.025em b}\kern-.08em
    T\kern-.1667em\lower.7ex\hbox{E}\kern-.125emX}}

\begin{document}

\title{A Survey of Transaction Tracing Techniques for Blockchain Systems}

%

%
\author{\IEEEauthorblockN{Ayush Kumar and Vrizlynn L.L. Thing} \\
\IEEEauthorblockA{Cyber Security Strategic Technology Centre \\
ST Engineering \\
}}
%
%

\maketitle

\begin{abstract}
With the proliferation of new blockchain-based cryptocurrencies/assets and platforms that make it possible to transact across them, it becomes important to consider not just whether the transfer of coins/assets can be tracked within their respective transaction ledger, but also if they can be tracked as they move across ledgers. This is especially important given that there are documented cases of criminals attempting to use these cross-ledger trades to obscure the flow of their coins/assets. 
In this paper, we perform a systematic review of the various tracing techniques for blockchain transactions proposed in literature, categorize them using multiple criteria (such as tracing approach and targeted objective) and compare them. Based on the above categorization, we provide insights on the state of blockchain transaction tracing literature and identify the limitations of existing approaches. Finally, we suggest directions for future research in this area based on our analysis.
\end{abstract}



\section{Introduction}
\label{intro}

Blockchain operates as a distributed ledger that immutably and transparently logs every party’s transactions. Its foundational traits- open visibility and the absence of a central authority have driven its uptake in industries such as supply chain management, finance, and the Internet-of-Things. In cryptocurrency, this technology enables users to mint and trade digital assets directly over peer-to-peer networks without intermediaries. Bitcoin, the foremost digital currency, now exceeds a \$2.3 trillion market valuation. While cryptocurrency has facilitated legitimate financial activity, its pseudo-anonymity has also attracted bad actors who launder money, commit fraud, and orchestrate hacks and scams. A Chainalysis report \cite{chainalysis-rep} estimates that illicit cryptocurrency-related transactions resulted in over \$40.9 billion in losses during 2024. As a result, extensive efforts concentrate on tracing and de-anonymizing suspect funds to disrupt criminal flows and help victims recover stolen assets.

Non-custodial exchange services such as Multichain\footnote{https://www.multichain.com/} and Synapse\footnote{https://synapseprotocol.com/} have recently proliferated, enabling users to swap tokens without relinquishing custody to any intermediary. This movement reflects the explosive growth in cryptocurrency offerings- CoinMarketCap registered only 36 cryptocurrencies in September 2013 with just seven exceeding a \$1 million market cap whereas by January 2025 that figure had swelled to about 16.29 million, and the ten largest collectively held over \$3.028 trillion in value\footnote{https://coinmarketcap.com/historical/20250105/}. In light of both the sheer volume of new coins and the rise of seamless cross-chain swaps, it is crucial to monitor fund flows not just within single ledgers but also as assets move between different blockchains. Such scrutiny is especially important given documented instances of criminals exploiting these cross-chain swaps to mask illicit proceeds.


Since May 2021, DeFi bridge breaches have inflicted over \$3.1 billion in cumulative losses, with new exploits occurring relentlessly. Immunefi \cite{immunefi} reports that white-hat researchers have been awarded more than \$20 million through bug bounty programs, averting nearly \$1 billion in further damages. Cross-chain bridge attacks now occupy the top spots on the DeFi incident leaderboard \cite{theblock, defi-liyi, rekt}, cementing them as prime targets for cybercriminals. The current environment remains dire, marked by a steady stream of hacks \cite{cteam, reynolds}. We posit that the complex interconnection of multiple blockchains has amplified these protocols’ risk exposure. The substantial funds locked within them serve as a tempting bait for attackers.


\subsection{Research Questions and Contributions}
In this paper, we systematize knowledge about the tracing techniques for blockchain transactions. So far, this information has been scattered among multiple sources. To achieve this goal, in Section \ref{background}, we first provide the relevant background knowledge on same-chain (occuring in a single blockchain) and cross-chain transactions necessary to understand this work. Then, we provide two contributions answering a specific research question (RQ), pointing to a relevant section discussing it.

\textit{RQ1: What are the different types of approaches proposed for tracing blockchain transactions? What are different targeted objectives for tracing? What are the various ways in which those tracing techniques handle false positives/negatives? What are the different types of transactions identified?} We analyze the existing blockchain tracing approaches based on above RQ and categorize them wherever possible. The classified papers represent works published till March 2025, underscoring the timeliness of this work. Section \ref{taxonomy} answers this RQ.


\textit{RQ2: Based on the existing gaps, what are potential best practices and avenues for future research to improve blockchain transaction tracing?} Our in-depth review of the current literature distills a set of proven recommendations and uncovers promising avenues for future inquiry. These findings are tailored to give blockchain tracing architects, developers, and analysts a solid foundation upon which to build and refine their work. Section \ref{future-dir} delivers the response to this RQ. Finally, Section \ref{literature} places our contributions in the context of existing scholarship, and Section \ref{conclusion} offers our closing reflections and projections for what lies ahead.


\subsection{Research Methodology}
We conduct a systematic literature review by crawling conference/journal/pre-print papers and theses using Google Scholar’s keyword search and forward reference search. The search was limited to papers since 2015 due to the limited amount of research available before that period. Next, we create a comprehensive database of all the documents found including useful attributes such as publication venue and date, proposed tracing techniques, approach to handle false positives/negatives, targeted types of transactions, trading platforms used, findings of evaluation on real-world blockchains, capabilities and limitations and details of opens-source implementations. We use the above database to systematize knowledge in the field by categorizing the tracing techniques proposed and comparing their capabilities and limitations. Based on our analysis, we derive insights, identify gaps and propose future directions for research into blockchain transaction tracing.

\section{Background}
\label{background}

\subsection{Types of blockchain transactions}
In this section, we present a primer on the various types of blockchain transactions since we are interested in studying techniques for tracing them.

\subsubsection{Cryptocurrency Transactions}
We briefly explain the working of cryptocurrency transactions on a blockchain. Assume that the first party, Alice wants send cryptocurrency coins to the second party, Bob, both have accounts with addresses. Alice broadcasts a message with the number of coins to be transferred and Bob’s address signed by her to the underlying blockchain network. The blockchain nodes verify the transaction and start mining to record the transaction on the blockchain until one of the nodes is selected. Once there is a consensus on the updated block being legitimate, the transaction gets recorded and Bob receives the coins in his account.

\subsubsection{Smart Contracts}

Smart contracts \cite{smart-contract} are self-executing pieces of code deployed on blockchain networks, enabling automated interactions among distrustful parties without a centralized intermediary. Because of the immutable nature of the blockchain, once deployed, smart contracts' logic cannot be altered even by their creators. This trustworthiness has fuelled the rapid emergence of sectors like Decentralized Finance (DeFi) and NFT marketplaces. Ethereum pioneered smart contracts with its Ethereum Virtual Machine (EVM), supporting two kinds of accounts: externally owned accounts (EOAs) managed via private keys, and contract accounts governed by on-chain code. Accordingly, transactions are split into external ones initiated by EOAs and internal ones triggered by contracts themselves.

\subsubsection{Same-chain Transactions}

Executing a transaction on a distributed ledger updates its internal state and generates event logs corresponding to the actions taken. We call ``same-chain'' events the sets of attributes drawn from those on-chain logs, and we describe ``same-chain'' transactions as sequences comprised of multiple such events.

\subsubsection{Cross-chain Transactions}

As blockchain networks continue to proliferate, ensuring that they can interoperate has become crucial. Interoperability among chains enables seamless transfer of digital assets and information across distinct platforms and relies on an interoperability mechanism (IM). Here, we examine the distributed ledgers labeled {$l_1, l_2, ..., l_k$}. The IM’s architecture can be tailored to different requirements, for example, operating under centralized or decentralized control, or adopting distributed versus non-distributed structures.


The surveys \cite{smartsync, belchior} delineate interoperability into three distinct categories: \emph{asset exchange}, \emph{data migration}, and \emph{asset transfer}. Our study addresses the latter. Asset transfer protocols function by immobilizing (locking) or eliminating (burning) tokens on the source ledger, followed by generating (minting) matching tokens on the destination ledger \cite{survey-belchior}. Once tokens are escrowed on the originating chain whether via a centralized custodian, a multi-signature scheme, or a smart contract, the receiving chain undertakes validation. Such verification may be achieved by replicating the source chain’s consensus mechanism within the target environment \cite{ethrelay, oana} or by employing zero-knowledge proofs \cite{harmonia, zkbridge}.


To formalize an asset movement between blockchains, the cross-chain transaction (\textit{cctx}) construct has been proposed. A \textit{cctx} comprises a series of cross-chain events that record state transitions across multiple ledgers. Each event carries two categories of attributes: native and extended (non-native) \cite{hephaestus}. Native attributes originate from on-chain events within their respective domains. Non-native attributes consist of extra metadata relevant only in cross-chain scenarios, examples include ledger identifiers, synchronized timestamps, token pricing information, or other off-chain data. This metadata is anchored on the chain via decentralized oracle networks, and its trustworthiness hinges on the oracle network’s integrity and the cross-chain protocol agreements.  

\begin{figure}[h]
\centering
\includegraphics[scale=0.35]{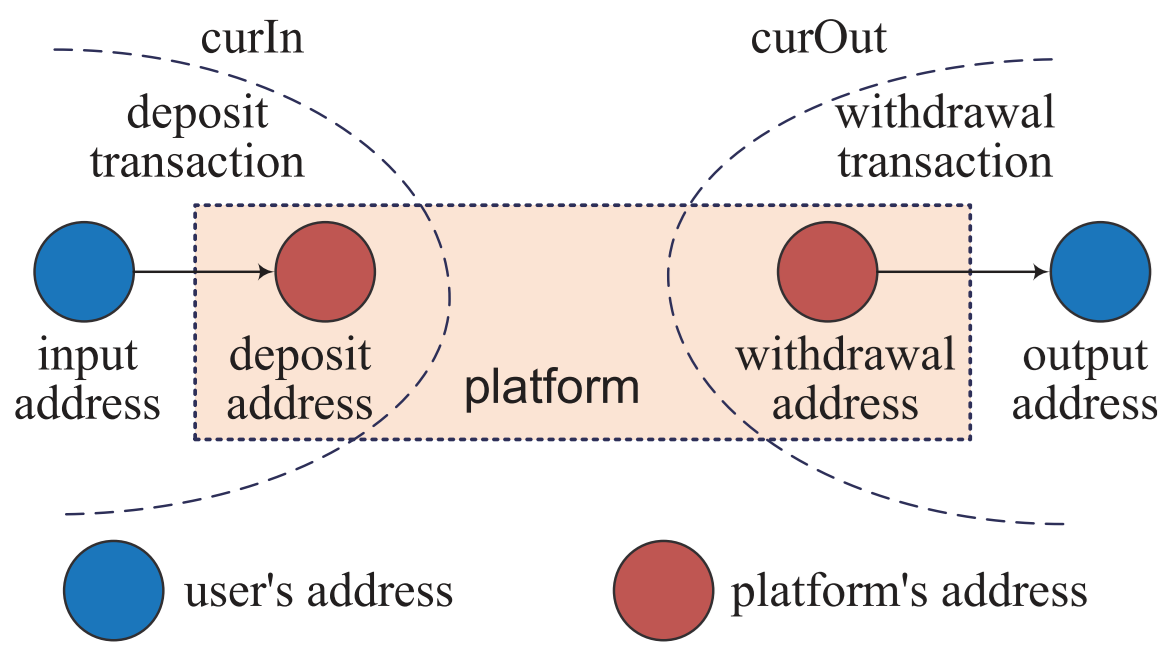}
\caption{The cross-ledger transaction flow model \cite{cltracer}.}
\label{cross-chain-tx}
\end{figure}

In a cross-ledger exchange workflow (see Fig. \ref{cross-chain-tx}), two distinct transactions occur: an initial deposit followed by a subsequent withdrawal. We refer to the source asset as $\mathrm{curIn}$ and the target asset as $\mathrm{curOut}$. To initiate an exchange, a user identifies the desired $\mathrm{curIn}$/$\mathrm{curOut}$ pair and provides a destination address for the $\mathrm{curOut}$ funds. The bridge platform then issues both an exchange ratio and a deposit address for $\mathrm{curIn}$. The user completes the deposit by transmitting the chosen amount of $\mathrm{curIn}$ to this deposit endpoint. After confirmation of this on-chain transfer, the bridge platform disburses the corresponding $\mathrm{curOut}$ quantity minus any applicable fees to the user’s specified withdrawal address.

\section{Blockchain Transaction Tracing Taxonomy}
\label{taxonomy}
In this section, we present a taxonomy of the blockchain transaction tracing techniques in literature.
\subsection{Tracing Approaches}
Transaction tracing techniques can be classified into \textit{heuristic-based}, \textit{rule-based} or \textit{graph learning-based} depending on their approach to identify related transactions.
\subsubsection{Heuristic-based}
These techniques use ad hoc algorithms based on parameters such as transaction time, blockchain block height and cross-chain bridge API for validation \cite{usenix19}. Some works are based on address relationships, for example, they exploit a central deposit address \cite{cltracer} or exchange deposit/hot wallet address \cite{evonax}. Techniques that target smart contract-based transactions may use features extracted from transaction traces and execution logs of bridge contracts \cite{connector, abctracer}.

\subsubsection{Rule-based} 
Such techniques build rules for detecting anomalous transactions based on an observation of characteristics of prior normal transactions. For example, \cite{jigsaw} inspects unsupported tokens, chronologically sorts all related transfer events, builds a fund flow graph and compares sent and released token values on source/target chains.

\subsubsection{Graph learning-based}
A number of works have use machine/deep learning methods on graph-based representations of cross-chain networks to detect abnormal transactions. \cite{bsci24} constructs graphs from transaction data, extracts features using \textit{node2vec} and classifies abnormal nodes in the graph by integrating graph embedding learning and multi-model fusion (LR-XGBoost-GCN). 
In \cite{bridgeguard}, execution logs and trace data are converted into cross-chain transaction execution graphs (xTEGs). The detection process then unfolds in two phases: a global analysis that employs the \textit{Graph2vec} embedding algorithm to extract overarching structural features from each xTEG, followed by a local analysis that tallies the occurrences of sixteen specific network motifs to spot anomalous transaction patterns.

XSema \cite{xsema} collects on-chain transfer records via RPC calls to blockchain nodes and maps them into a directed asset-flow graph. It then derives topological descriptors by tallying the frequency of sixteen distinct network motifs. At the same time, it concatenates the sequence of event identifiers emitted by each transaction into a single text string. This string is passed through a pre-trained CodeBERT encoder, with its output embeddings fine-tuned using a multilayer perceptron. Finally, the model merges the structural motif vectors and the refined semantic embeddings into a unified feature vector for classification. 

In \cite{txgraph}, the authors introduce \textit{TxGraph}, a graph-based model that encapsulates the relationships between flagged accounts and their neighboring nodes. They apply a custom augmentation procedure to mimic the development of illicit transaction patterns, producing positive sample pairs. These pairs are then fed into a graph contrastive learning framework to derive distinctive feature embeddings from unlabeled blockchain data. 
In \cite{tpgraph}, blockchain tracing is approached as a set of temporal graph queries over a graph-structured ledger that has been split into discrete time intervals. The entire blockchain history is divided into these chronological slices, with each slice represented by its own sub-graph for trace analysis. A hierarchical graph framework is then erected to capture dependencies between these sub-graphs across time. Tracing tasks execute in parallel on all sub-graphs, and their outputs are either combined to seed the next iteration or aggregated as the final tracing result.

\subsection{Targeted Objective}
We have categorized the transaction tracing techniques in literature based on their objective that they seek to achieve.
\subsubsection{Insights from transaction analysis}
This objective focuses on usage of trading platforms, and the potential they offer in terms of the ability to track flows of coins as they move across different transaction ledgers. Using various heuristics/rule-based/learning-based techniques for tracing, such works also offer insights on types of cross-chain transactions, clustering analysis of transactions/addresses. Transactions can be in terms of coins or executed using smart contracts.

\subsubsection{Tracing detection performance}
Some works embed transaction data in graphs, extract features and use ML/GNN models to detect abnormal transactions. Their aim is to achieve high detection performance than benchmark tracing techniques in terms of metrics such as precision/recall/F1 scores.

\subsubsection{Tracing query speedup}
Graph-based transaction tracing often overlooks the temporal dimension when transforming blockchain records into a graph. Trace analyses usually target a limited set of transactions confined to a user-defined time interval, meaning only the edges for those transactions are needed. Ignoring this time-based constraint during graph construction forces traversal algorithms to process many irrelevant edges, wasting resources. For instance, \cite{tpgraph} addresses this by pruning the blockchain graph to include only the edges that fall within the query’s time window, thus cutting down the overhead of breadth-first searches over a specified period.

\subsection{FP/FN Handling Approach}
There are bound to be false positives (identifying a transaction when there is not one) and false negatives (missing legitimate transactions) while applying transaction tracing techniques, which are both not productive as the former leads to an increase in tracing effort while the latter creates a situation where the tracing effort is negated due to missing links. We have categorized the approaches taken by tracing techniques while handling false positives and false negatives into \textit{API-based} and \textit{Address-based}.
\subsubsection{API-based}
In \cite{usenix19}, the authors verify candidate ShapeShift transactions by sending recipient addresses to the ShapeShift API, confirming that the API recognizes the address and that the outgoing currency matches expectations.  Occasionally, when users re-use deposit addresses and the API returns only the most recent transaction for that address, the response may correspond to a different transfer than intended, causing false positives.  To eliminate these mismatches, they require the API’s transaction record to concur with the target ShapeShift exchange on three criteria: the specific currency pair, the exact transferred amount, and a timestamp falling within the paper’s defined window.
To reduce false negatives, the authors have recommended to increase the searched block radius with respect to the underlying blockchain for both the transaction detection heuristics proposed. While analysing trading activity of bots operating on ShapeShift, the authors obtained 107 trading clusters. Transactions which were believed to be false positives were identified by assuming that they all have a similar value but were outliers.

\subsubsection{Address-based}
In \cite{cltracer}, the authors tackle clustering false positives by first isolating input addresses with high historical send volumes, gradually reducing the volume threshold before running their combined cross-ledger clustering (CCLC) heuristic. Then, to uncover output addresses associated with public services that might distort the largest cluster, they record how often each address appears within that cluster and sort the addresses in descending order based on these counts.

\subsection{Transaction Types Identified} We have categorized the various types of cryptocurrency transactions identified from literature based on an analysis of historical transactions collected from cross-chain bridge platforms.

\begin{itemize}
    \item \textit{Pass-through}: End-to-end asset movements where funds are deposited on one blockchain and withdrawn on another.
    \item \textit{U-turn}: Transfers in which coins sent to a secondary chain are almost immediately routed back to their chain of origin.
    \item \textit{Round-trip}: A fusion of pass-through and U-turn activities. Assets travel from the original chain to a second one and then return, leaving the user with fewer tokens of the starting currency due to fees, exchange rates, and gas costs.
    \item \textit{Coin swap}: Same-currency conversions executed on exchanges. Despite resulting in a smaller quantity of the identical token, this approach conceals the true sender–recipient path by sourcing payouts from other users’ deposits or the exchange’s hot wallet rather than directly forwarding from user's deposit address to the withdrawal address.
\end{itemize}


\subsection{Targeted Malicious Transaction Types}
Over the past few years, the evolution of blockchain platforms has propelled cryptocurrencies, particularly Bitcoin and Ethereum, into the spotlight. By July 2025, their combined market capitalization surpassed \$3.78 trillion. The distributed ledger architecture and transaction pseudonymity amplify the potential for misuse. 
We have categorized the malicious activities targeted by the transaction tracing techniques covered in this survey.
\subsubsection{Money laundering}
Money laundering refers to the illicit conversion of proceeds from criminal enterprises, such as narcotics trafficking or terrorist financing, into assets that appear legitimate, thereby concealing their origin. This scheme generally unfolds in three phases: (i) \textit{Placement}, where illicit cash is injected into the financial system, often by splitting large sums into smaller deposits across numerous accounts to bypass AML (Anti-Money Laundering) monitoring; (ii) \textit{Layering}, involving a web of complex, high-frequency transactions designed to detach funds from their illegal source and thwart tracing efforts; and (iii) \textit{Integration}, the stage at which cleansed capital is reintroduced into the economy as apparently lawful resources. Money laundering in blockchains, especially the layering phase, uses a variety of transaction types such as multiple pass-through transactions (which transfer cryptocurrency from one blockchain to another), round-trip transactions (which transfer cryptocurrency from one blockchain to another and then back) and coin swaps (exchanging of funds within same cryptocurrency) to obfuscate the origin of funds. 

\subsubsection{Phishing scam}
When an attacker seizes private keys or other access credentials, they initiate unauthorized fund movements. These illicit withdrawals, though rare, usually involve large sums and occur soon after the breach. The compromised account then redirects the stolen assets to addresses under the fraudster’s control.

\subsubsection{Rug pull}
%
Rug pull schemes begin when an operator mints a coin lacking real utility and lists it on a decentralized trading platform, pairing it with a major cryptocurrency to seed initial liquidity. They then launch targeted promotions across social media to generate hype and coax users into swapping stablecoins for the new token. Once sufficient conversions have occurred, the fraudster drains the liquidity reserves, absconds with the paired assets, and abandons investors with valueless tokens. These operations frequently combine with rapid market manipulation tactics, most notably pump-and-dump manoeuvres, that artificially drive up the token’s price and trading volume in a short span. This contrived surge creates a sense of urgency among novice traders, who buy in at inflated levels. At the peak of this frenzy, scammers either liquidate their positions at the heightened prices or completely purge the pool, reclaiming their original stake plus any excess proceeds, while victims are left holding depreciated tokens.


\subsubsection{Darknet}
In darknet markets, buyers’ payments are routed through a sequence of intermediary addresses to conceal their provenance before arriving at vendors. These transfers take place sporadically, driven by the availability of illicit goods and shifting demand dynamics. Transaction volumes can spike or wane in response to user interest and law enforcement pressure. The ecosystem consists of pseudonymous purchasers, underground sellers, and covert platform administrators. Darknet markets usually involve multiple pass-through transactions.

\subsubsection{Ponzi Schemes}
In a Ponzi arrangement, participants are lured by promises of large gains with virtually no risk, yet the scheme survives solely by reallocating funds from later investors to satisfy earlier ones. This cycle perpetuates the appearance of genuine returns while relying on a constant influx of new capital. Transaction patterns for these scams usually exhibit waves of incoming investments immediately followed by regular disbursements to the scheme’s initial backers.  

\subsection{Dataset Curation}
We have identified different approaches to creating the datasets used for training and testing the transaction tracing methods covered in this survey. Works on same-chain transaction tracing \cite{earldet, t-racer} download full node data corresponding to a blockchain (e.g., BTC, ETH) using custom tools (e.g., Bitcoin Core) till or after a certain block, parse and process the downloaded data. Ground truth labels for malicious activities are obtained based on information available from attack case studies published by blockchain security companies, public forums, datasets and previous papers. 

Works on cross-chain transaction tracing \cite{crossaad, bsci24, jigsaw} download the raw transaction data from cross-chain bridge platforms by using developer APIs or simply crawling/parsing their web explorer URLs, and subsequently label and process the downloaded data. To obtain the ground truth labels, either the Etherscan or the bridge explorer is used to identify abnormal addresses (exploit addresses, rug pull addresses, addresses associated with phishing scammers).

Works on cross-chain transaction tracing using bridge smart contracts \cite{connector, abctracer, bridgeguard} collect contract and transaction data for the selected bridges through the bridges' official websites and construct the ground truth deposit-withdrawal transactions using the bridge explorers. Cross-chain bridge attacks are identified from academic papers and incident summaries by security companies. Subsequently, logs and other transaction-related data on the attacks are obtained using custom tools such as BlockchainSpider. 

\subsection{Clustering Analysis Results}
Some works perform clustering analysis to analyze the relationships between addresses used in cryptocurrency transactions. This may help detect illicit behaviors, especially when combined with existing single-ledger clustering techniques.

In \cite{usenix19}, the researchers apply a \textit{shared ownership heuristic} for clustering, which assumes that every input in a multi-input transaction is controlled by the same entity. Their analysis showed that exchanges maintain separate clusters for each cryptocurrency and even adapt their clustering behavior per asset. Because this approach sometimes merged user-controlled wallets with exchange clusters, the authors proposed a \textit{common relationship heuristic} instead: any two addresses (potentially on different chains) that send to or receive from the same address are likely linked socially. Deploying this alternative revealed two key findings: (1) when clusters are ranked by in-degree, ShapeShift stands out as the predominant platform among CoinPayments.net users, and (2) ordering by out-degree uncovers a substantial number of users preferring ShapeShift for currency swaps rather than traditional exchanges.  

In \cite{cltracer}, the authors employ the Combined Cross-Ledger Clustering (CCLC) heuristic, which infers that whenever two input addresses’ output address sets intersect, all addresses within those sets share a single owner. Upon ordering clusters by descending total ETH deposits, they discovered that (1) most clusters comprised just one input address, and (2) the largest cluster was driven by output addresses belonging to public services.

\begin{table*}
	\centering
	\begin{threeparttable}
	\caption{Classification of Cryptocurrency Transaction Tracing Papers}
    \label{det-perf-table}
    \begin{tabular}{ | c | c | c | c | c | c | c | c | c | c | c | c | }
    \hline
    {\textbf{Ref}} & {\textbf{Year}} & {\textbf{Technique}} & \multicolumn{3}{|c|}{\textbf{Tracing approach}} & \multicolumn{3}{|c|}{\textbf{Target obj.}} & \multicolumn{3}{|c|}{\textbf{Target Tx type}} \\ \cline{4-12}
	& & & Heu & Rul & Gbl & Ins & Det & Qsp & Cnmt & Cmt & Smt \\ \hline
    \cite{usenix19} & 2019 & {Uses transaction time and block height} & \Checkmark & \XSolidBrush & \XSolidBrush & \Checkmark & \XSolidBrush & \XSolidBrush & \Checkmark & \XSolidBrush & \XSolidBrush \\ \hline
    \cite{cltracer} & 2022 & {Exploits a central deposit address} & \Checkmark & \XSolidBrush & \XSolidBrush & \Checkmark & \XSolidBrush & \XSolidBrush & \Checkmark & \XSolidBrush & \XSolidBrush \\ \hline
    \cite{connector} & 2024 & \makecell{Uses transaction traces and execution\\ logs of bridge contracts} & \Checkmark$^{*}$ & \XSolidBrush & \XSolidBrush & \Checkmark & \XSolidBrush & \XSolidBrush & \Checkmark & \XSolidBrush & \XSolidBrush \\ \hline
    \cite{abctracer} & 2025 & \makecell{Improves upon \cite{connector} and uses NER\\ to learn cross-chain clues automatically} & \XSolidBrush & \XSolidBrush & \Checkmark & \XSolidBrush & \Checkmark & \XSolidBrush & \Checkmark & \XSolidBrush & \XSolidBrush \\ \hline
    \cite{bsci24} & 2024 & \makecell{Integrates graph embedding learning and\\ multi-model fusion} & \XSolidBrush & \XSolidBrush & \Checkmark & \XSolidBrush & \Checkmark & \XSolidBrush & \XSolidBrush & \XSolidBrush & \Checkmark \\ \hline
    \cite{jigsaw} & 2024 & \makecell{Compares sent and released token values\\ on source/target chains} & \XSolidBrush & \Checkmark & \XSolidBrush & \Checkmark & \XSolidBrush & \XSolidBrush & \XSolidBrush & \Checkmark & \XSolidBrush \\ \hline
    \cite{bridgeguard} & 2025 & {Uses global and local graph mining} & \XSolidBrush & \XSolidBrush & \Checkmark & \Checkmark & \XSolidBrush & \XSolidBrush & \XSolidBrush & \Checkmark & \XSolidBrush \\ \hline
    \cite{xsema} & 2024 & \makecell{Classifies concatenated semantic representation\\ of asset transfer and message-passing text\\ extracted from event logs} & \XSolidBrush & \XSolidBrush & \Checkmark & \XSolidBrush & \Checkmark & \XSolidBrush & \Checkmark & \XSolidBrush & \XSolidBrush \\ \hline
    \cite{crossaad} & 2024 & {Based on XGBoost classifier} &\XSolidBrush  & \XSolidBrush & \Checkmark & \XSolidBrush & \Checkmark & \XSolidBrush & \XSolidBrush & \Checkmark & \XSolidBrush \\ \hline
    \cite{txgraph} & 2024 & \makecell{Uses graph contrastive mechanism to learn\\ discriminative features from unlabeled data} & \XSolidBrush & \XSolidBrush & \Checkmark & \XSolidBrush & \Checkmark & \XSolidBrush & \XSolidBrush & \XSolidBrush & \Checkmark \\ \hline
    \cite{tpgraph} & 2024 & \makecell{Conducts tracing to time-partitioned sub-graphs\\ in parallel and merges tracing results } & \XSolidBrush & \XSolidBrush & \XSolidBrush & \XSolidBrush & \XSolidBrush & \Checkmark & \XSolidBrush & \XSolidBrush & \XSolidBrush \\ \hline
    \cite{evonax} & 2025 & \makecell{Uses transaction time, exchange form\\ and exchange deposit/hot wallet address} & \Checkmark & \XSolidBrush & \XSolidBrush & \Checkmark & \XSolidBrush & \XSolidBrush & \XSolidBrush & \Checkmark & \XSolidBrush \\ \hline
    \cite{monero}$^{**}$ & 2019 & {Exploits information leaked by currency hard forks} & \Checkmark & \XSolidBrush & \XSolidBrush & \Checkmark & \XSolidBrush & \XSolidBrush & \XSolidBrush & \XSolidBrush & \Checkmark$^{***}$ \\ \hline
    \cite{monero17}$^{**}$ & 2017 & \makecell{Leverages network and usage statistics to\\ propose heuristics for traceability attacks} & \Checkmark & \XSolidBrush & \XSolidBrush & \Checkmark & \XSolidBrush & \XSolidBrush & \XSolidBrush & \XSolidBrush & \Checkmark$^{***}$ \\ \hline
    \cite{ethheist} & 2024 & \makecell{Uses algorithm based on augmented poison policy} & \Checkmark & \XSolidBrush & \XSolidBrush & \Checkmark & \XSolidBrush & \XSolidBrush & \XSolidBrush & \XSolidBrush & \Checkmark \\ \hline
    \cite{t-racer} & 2023 & \makecell{Uses graph expansion and local community\\ detection based on a ranking algorithm} & \XSolidBrush & \XSolidBrush & \Checkmark & \XSolidBrush & \Checkmark & \XSolidBrush & \XSolidBrush & \XSolidBrush & \Checkmark \\ \hline
    \cite{earldet} & 2025 & \makecell{Uses clustering-based path selector to\\ find sibling addresses and GCN module\\ for capturing path inter-relation} & \XSolidBrush & \XSolidBrush & \Checkmark & \XSolidBrush & \Checkmark & \XSolidBrush & \XSolidBrush & \XSolidBrush & \Checkmark \\ \hline
    \end{tabular}
    \begin{tablenotes}
	\item \Checkmark- Satisfies criteria
	\item \XSolidBrush- Does not satisfy criteria
	\item ${}^*$Uses heuristic (during withdrawal transaction matching) and ML techniques (during deposit transaction identification), heuristic being final step.
	\item ${}^{**}$Monero is a privacy-focused chain with inbuilt mechanisms such as mixin and ringCT to enhance privacy and prevent tracing.
	\item ${}^{***}$Target transaction type is normal and malicious transactions (in terms of non-trivial rings \cite{monero}) on the same chain.
	\end{tablenotes}
	\end{threeparttable}
\end{table*}
  
\begin{table*}
	\centering
	\begin{threeparttable}
	\caption{Classification of Cryptocurrency Transaction Tracing Papers (part 2)}
    \label{det-perf-table-2}
	\vspace*{0.15in}  
    \begin{tabular}{ | c | c | c | c | c | c | c | c | c | c | c | c | c | c | }
    \hline
    {\textbf{Ref}} & \multicolumn{2}{|c|}{\textbf{Target Entity}} & \multicolumn{2}{|c|}{\textbf{\thead{Target no.\\ of platforms}}} & \multicolumn{2}{|c|}{\textbf{\thead{Same-chain OR\\ Cross-chain}}} & \multicolumn{2}{|c|}{\textbf{FP/FN handling}} & \multicolumn{5}{|c|}{\textbf{Target Mal. tx. type}$^{\#\#}$} \\ \cline{2-14}
	& Cry & Sct & Spm & Mpm & Sch & Cch & Api & Add & Ph & ML & Pon & DN & RP \\ \hline
    \cite{usenix19} & \Checkmark & \XSolidBrush & \Checkmark & \XSolidBrush & \XSolidBrush & \Checkmark & \Checkmark & \XSolidBrush & \XSolidBrush & \XSolidBrush & \XSolidBrush & \XSolidBrush & \XSolidBrush \\ \hline
    \cite{cltracer} & \Checkmark & \XSolidBrush & \XSolidBrush & \Checkmark & \XSolidBrush & \Checkmark & \XSolidBrush & \Checkmark & \XSolidBrush & \XSolidBrush & \XSolidBrush & \XSolidBrush & \XSolidBrush \\ \hline
    \cite{connector} & \XSolidBrush & \Checkmark & \XSolidBrush & \Checkmark & \XSolidBrush & \Checkmark & \XSolidBrush$^{\#}$ & \XSolidBrush & \XSolidBrush & \XSolidBrush & \XSolidBrush & \XSolidBrush & \XSolidBrush \\ \hline
    \cite{abctracer} & \XSolidBrush & \Checkmark & \XSolidBrush & \Checkmark & \XSolidBrush & \Checkmark & \XSolidBrush$^{\#}$ & \XSolidBrush & \XSolidBrush & \XSolidBrush & \XSolidBrush & \XSolidBrush & \XSolidBrush \\ \hline
    \cite{bsci24} & \Checkmark & \XSolidBrush & \Checkmark & \XSolidBrush & \XSolidBrush & \Checkmark & \XSolidBrush$^{\#}$ & \XSolidBrush & \XSolidBrush & \XSolidBrush & \XSolidBrush & \XSolidBrush & \XSolidBrush \\ \hline
    \cite{jigsaw} & \Checkmark & \XSolidBrush & \XSolidBrush & \Checkmark & \XSolidBrush & \Checkmark & \XSolidBrush$^{\#}$ & \XSolidBrush & \XSolidBrush & \XSolidBrush & \XSolidBrush & \XSolidBrush & \XSolidBrush \\ \hline
    \cite{bridgeguard} & \Checkmark & \XSolidBrush & \XSolidBrush & \Checkmark & \XSolidBrush & \Checkmark & \XSolidBrush$^{\#}$ & \XSolidBrush & \XSolidBrush & \XSolidBrush & \XSolidBrush & \XSolidBrush & \XSolidBrush \\ \hline
    \cite{xsema} & \Checkmark & \XSolidBrush & \XSolidBrush & \Checkmark & \XSolidBrush & \Checkmark & \XSolidBrush$^{\#}$ & \XSolidBrush & \XSolidBrush & \XSolidBrush & \XSolidBrush & \XSolidBrush & \XSolidBrush \\ \hline
    \cite{crossaad} & \Checkmark & \XSolidBrush & \Checkmark & \XSolidBrush & \XSolidBrush & \Checkmark & \XSolidBrush$^{\#}$ & \XSolidBrush & \XSolidBrush & \XSolidBrush & \XSolidBrush & \XSolidBrush & \XSolidBrush \\ \hline
    \cite{txgraph} & \Checkmark & \XSolidBrush & \XSolidBrush & \Checkmark & \Checkmark & \XSolidBrush & \XSolidBrush$^{\#}$ & \XSolidBrush & \Checkmark & \Checkmark & \Checkmark & \Checkmark & \XSolidBrush \\ \hline
    \cite{tpgraph} & \Checkmark & \XSolidBrush & \XSolidBrush & \Checkmark & \Checkmark & \XSolidBrush & \XSolidBrush$^{\#}$ & \XSolidBrush & \XSolidBrush & \XSolidBrush & \XSolidBrush & \XSolidBrush & \XSolidBrush \\ \hline
    \cite{evonax} & \Checkmark & \XSolidBrush & \Checkmark & \XSolidBrush & \Checkmark & \XSolidBrush & \XSolidBrush$^{\#}$ & \XSolidBrush & \XSolidBrush & \Checkmark & \XSolidBrush & \XSolidBrush & \Checkmark \\ \hline
    \cite{monero} & \Checkmark & \XSolidBrush & \Checkmark & \XSolidBrush & \Checkmark & \XSolidBrush & \XSolidBrush$^{\#}$ & \XSolidBrush & \XSolidBrush & \XSolidBrush & \XSolidBrush & \XSolidBrush & \XSolidBrush \\ \hline
    \cite{monero17} & \Checkmark & \XSolidBrush & \Checkmark & \XSolidBrush & \Checkmark & \XSolidBrush & \XSolidBrush$^{\#}$ & \XSolidBrush & \XSolidBrush & \XSolidBrush & \XSolidBrush & \XSolidBrush & \XSolidBrush \\ \hline
    \cite{ethheist} & \Checkmark & \XSolidBrush & \Checkmark & \XSolidBrush & \Checkmark & \XSolidBrush & \XSolidBrush$^{\#}$ & \XSolidBrush & \XSolidBrush & \Checkmark & \XSolidBrush & \XSolidBrush & \XSolidBrush \\ \hline
    \cite{t-racer} & \Checkmark & \XSolidBrush & \XSolidBrush & \Checkmark & \Checkmark & \XSolidBrush & \XSolidBrush$^{\#}$ & \XSolidBrush & \XSolidBrush & \XSolidBrush & \XSolidBrush & \XSolidBrush & \Checkmark \\ \hline
    \cite{earldet} & \Checkmark & \XSolidBrush & \Checkmark & \XSolidBrush & \Checkmark & \XSolidBrush & \XSolidBrush$^{\#}$ & \XSolidBrush & \XSolidBrush & \XSolidBrush & \XSolidBrush & \Checkmark & \XSolidBrush \\ \hline
    \end{tabular}
    \begin{tablenotes}
	\item ${}^{\#}$The paper does not mention the FP/FN handling method explicitly. Therefore, we cannot say for sure if it uses API-based or address-based methods.
	\item ${}^{\#\#}$Here, \XSolidBrush means that the paper does not mention the corresponding malicious transaction type explicitly.
	\end{tablenotes}
	\end{threeparttable}
\end{table*}

\section{Status of Blockchain Transaction Tracing}
In this section, we present the results of our work and extensively discuss the most relevant insights. We gather all relevant insights and propose guidelines for designing tracing techniques in future.

\subsection{Comparison Framework}
We classify 15 academic papers in light of the design and usability characteristics presented in the previous sections which are relevant to both tracing technique designers and platform users. The classification is presented in Tables \ref{det-perf-table} and \ref{det-perf-table-2}.


\subsubsection{Classification Criteria}
We present the criteria on which we base ourselves to classify the transaction tracing techniques proposed in the literature. 
\begin{itemize}
 \item Tracing approach- based on heuristic (Heu), rules (Rul) or machine/graph-based learning (Gbl)
 \item Targeted objective- deriving insights from tracing (Ins), improving tracing detection performance (Det) or speeding up tracing query (Qsp) 
 \item Targeted transaction type- A tracing technique may target both normal and malicious cross-chain transactions (Cnmt), malicious cross-chain transactions (Cmt) or malicious same-chain transactions (Smt). Here, normal transactions refer to legitimate transfer of coins/assets between transacting parties while malicious transactions refer to those which are part of attacks such as ransomware, rug pull, smart contract vulnerability exploitation.
 \item Targeted blockchain entity- A tracing technique may be designed for blockchains dealing with cryptocurrency (Cry) or smart contract (Sct).
 \item Single platform (Spm) vs multiple platforms (Mpm)- A tracing technique may be designed for a specific platform (e.g., Bitcoin, Ethereum, Shapeshift) or it may be platform agnostic. 
 \item Same-chain (Sch) vs Cross-chain (Cch)- A tracing technique may be designed to trace transactions on a single blockchain or transactions across ledgers.
 \item Targeted Malicious transaction type- A tracing technique may target phishing scam (Ph), money laundering (ML) or ponzi scheme (Pon) or darknet (DN) or rug-pull (RP) 
 \item FP/FN handling approach- API-based (Api) or address-based (Add)

\end{itemize}

\subsubsection{Insights}
We now present a list of insights taken from our analysis of the literature.
\begin{itemize}
	\item Almost an equal number of works are targeted towards ``deriving insights from tracing'' (53.33\%) and ``improving tracing detection performance'' (40\%). However, just one work \cite{tpgraph} is focused on ``speeding up tracing queries''.	
	\item An equal number of works propose tracing techniques based on heuristics (40\%) and graph-based learning (40\%). However, there is just one work \cite{jigsaw} which proposes a rule-based tracing technique.
	\item Most works are focused on ``analysing or detecting malicious same-chain transactions'' (40\%). An equal number of works are focused on ``analysing or detecting malicious cross-chain transactions'' (26.67\%) and ``analysis of both normal and malicious cross-chain transactions''(26.67\%).
	\item Most works target tracing of cryptocurrency coin transactions (93.33\%) while the rest target tracing of asset transactions executed by smart contracts.
	\item Almost an equal number of works are targeted at a single bridge/cryptocurrency platform (46.67\%) and multiple platforms (53.33\%).
	\item Almost an equal number of works are targeted at ``same-chain'' transactions (46.67\%) and cross-chain transactions (53.33\%).
	\item Money laundering is the illicit activity which is most targeted by transaction tracing techniques, followed by darknet and rug-pull while ponzi schemes are targeted the least.	 
	\item Very few works (13.33\%) provide details on how they handle false positives arising from their proposed transaction detection approach.
	\item All works on graph learning-based abnormal transaction detection use different datasets and benchmarks, making it difficult to compare performance across them. 
	\item \textit{Limitations of tracing techniques:} 
	\begin{enumerate}
		\item A few proposed techniques work for specific platforms/currencies only (\cite{usenix19} for Shapeshift, \cite{evonax} for Evonax, \cite{monero} for Monero).
		\item Heuristic algorithms depend on parameters such as deviation range of block height which need to be carefully selected.
		\item There are no open-source datasets (or working scripts to build datasets) released for blockchain transactions which creates a significant barrier for future researchers. \cite{usenix19} released crawler scripts which capture real-time transaction flows using Shapeshift API, however, the API does not work anymore.
		\item Techniques designed for smart contracts have a few common limitations: (1) transaction tracking for inconsistent deposit-withdrawal transactions and many-to-many cross-chain relationships are not in the scope, (2) they do not encompass non-fungible token (NFT) bridges, privacy enhancing bridges, or non-EVM-compatible bridges, and (3) their effectiveness largely depends on the availability of open-source contract codes. Bridges lacking these codes may pose challenges.

	\end{enumerate}
\end{itemize}


\subsection{Future Research Directions}
\label{future-dir}
We have identified several areas within the cryptocurrency tracing community that require further exploration and development. Firstly, there is a need to design a privacy-preserving cross-chain transaction scheme that preserves anonymity of transacting parties while ensuring some level of traceability. There has been little to no work done in this area so far and cross-chain privacy is predominantly maintained within underlying chains that inherently support privacy-enhancing features (e.g., Monero, Zerocash). Secondly, future works on cross-chain transaction racing should compare performance with multiple relevant SOTA techniques from this survey. As pointed out earlier, if all works use different datasets and benchmarks, it makes it difficult to compare performance across them. Thirdly, the trend of using graph learning instead of heuristics for transaction tracing should continue in order to address a wide variety of cryptocurrency/bridge platforms and abnormal transactions. Further, instead of using static graph representations, future works should also encode the temporal aspects of blockchain transactions to capture their dynamic nature. Finally, the research community needs to release open-source datasets of blockchain transactions or at least working scripts to build datasets and thus lower the entry barrier for future researchers.


\section{Related Work}
\label{literature}
Although scholars have conducted much research in this area in recent years, however, to the best of our knowledge, \textit{there is no existing survey focusing on blockchain transaction tracing}. Augusto et al. \cite{survey-augusto} have systematized the relevant security and privacy properties and approaches of blockchain interoperability solutions. From previous cross-chain bridge hacks and privacy leaks, they have identified 45 identified vulnerabilities and categorized them. Liang et al. \cite{survey-permission} have categorized the privacy of permissioned and permissionless blockchains and based on that, performed a systematic review of blockchain privacy protection literature. They have also highlighted the key technologies/approaches used for privacy protection. Qi et al. \cite{survey-graph-mine} have reviewed the literature on graph learning in blockchain data mining and categorized them using three data mining steps which are further sub-categorized. They have also summarized and compared public parsing tools and online platforms available for blockchain data mining. 

Hassan et al. \cite{survey-blockch-anomaly} have provided an overview of anomaly detection in blockchain technology. They have also classified anomalous attacks and their detection models, provided some fundamental metrics and key requirements for robust and timely identification of anomalies and highlighted critical challenges. Liu et al. \cite{survey-blocksys23} have reviewed the existing literature on detection of abnormal transactions in blockchains and categorized them as per data imbalance handling approach, feature extraction method, and classification algorithms used. Unlike our study, most research reviews do not cover cryptocurrency transaction tracing and cross-chain transactions, lack fine-grained criteria for classification of works and fail to provide any quantitative insights using that classification. Though \cite{survey-augusto} addresses cross-chain security and privacy properties and identifies multiple theoretical cross-chain system vulnerabilities, it never explores the tracing aspect of cross-chain transactions. We compare previous blockchain security, privacy and anomalous transaction detection studies with our work in Table \ref{survey-comp}.

\begin{table}[h]
	\centering
	\begin{threeparttable}
    \caption{Comparison with existing surveys}
    \label{survey-comp}
    \begin{tabular}{ | l | l | l | l | l | l | }
    \hline
    \textbf{Ref} & \textbf{Year} & \textbf{CT} & \textbf{TT} & \textbf{CC} & \textbf{QI} \\ \hline
    Our survey & 2025 & \Checkmark & \Checkmark & \Checkmark & \Checkmark \\ \hline
    \cite{survey-augusto} & 2024 & \Checkmark & \XSolidBrush & \Checkmark & \Checkmark \\ \hline
    \cite{survey-permission} & 2024 & \XSolidBrush & \XSolidBrush & \XSolidBrush & \XSolidBrush \\ \hline
    \cite{survey-graph-mine} & 2024 & \XSolidBrush & \XSolidBrush & \XSolidBrush & \XSolidBrush \\ \hline
    \cite{survey-blockch-anomaly} & 2022 & \XSolidBrush & \XSolidBrush & \Checkmark & \XSolidBrush \\ \hline
    \cite{survey-blocksys23} & 2024 & \XSolidBrush & \XSolidBrush & \XSolidBrush & \XSolidBrush \\ \hline
    \end{tabular}
    \begin{tablenotes}
	\item \textbf{CT}- Covers cross chain transactions
	\item \textbf{TT}- Covers transaction tracing
	\item \textbf{CC}- Identifies fine-grained classification criteria
	\item \textbf{QI}- Provides quantitative insights
	\end{tablenotes}
	\end{threeparttable}
\end{table}


%

\section{Conclusion}
\label{conclusion}
In this paper, we have performed a systematization of cryptocurrency/smart contract transaction tracing literature. We have categorized the tracing techniques proposed on basis of multiple criteria: tracing approach (heuristic/rule-based/graph learning-based), targeted objective (insights from analysis/tracing detection performance/tracing query speed-up). Furthermore, we have presented insights on the distribution of SOTA techniques as per above categories and their limitations. Our survey reveals a lack of research activity in tracing of smart contract-enforced transactions, limited details of false positives' handling in most works and prevalence of the practice of using different datasets and benchmarks in all works. Finally, we have proposed a few directions for future research in blockchain transaction tracing.


\bibliographystyle{ieeetran}
\begingroup
\raggedright
\bibliography{cryptotracebib}
\endgroup


\end{document}